\documentclass[12pt]{article}

\usepackage[dvips]{graphicx,color}

\begin{document}

\begin{center}

{\Large {\bf The Riches of the Elementary \\

\vspace{0,5cm}

Fluxbrane Solution}}

\vspace{2cm}

{Boris L. Altshuler}\footnote[1]{E-mail adresses: altshuler@mtu-net.ru \& altshul@lpi.ru}

\vspace{0,5cm}

{\it Theoretical Physics Department, P.N. Lebedev Physical
Institute, \\  53 Leninsky Prospect, Moscow, 119991, Russia}

\vspace{1,5cm}

{\bf Abstract}

\end{center}

The conventional approach to calculation of the radion effective potential 
in the string theory inspired models with magnetic fluxbrane throat-like 
space-time compactified on a sphere gives the analytical expressions hopefully 
capable to describe early inflation. Potential is rather flat inside the 
throat, possesses steep slope for reheating in vicinity of the top of the 
throat, and zero minimum at the top where UV brane's position is stabilized 
by the anisotropic junction conditions. The form of the effective radion 
potential is unambiguously determined by the choice of the theory. The D10
Type $IIA$ supergravity proves to be of 
special interest. In this theory the observed large value of the electro-weak 
hierarchy may be received. The Euclidian "time" version of the Schwarzshild 
type non-extremal generalization of the elementary fluxbrane solution is used
as a tool to fix the additional modulus - size of extra torus and to 
construct a smooth IR end of the throat; it also permits to estimate the 
small deviation of the radion effective potential from its zero value in 
the minimum which may be seen today as Dark Energy density. Thus most 
familiar fluxbrane solution proves to be rich enough in its possible 
physical predictions.

\vspace{0,5cm}

\newpage
\section{Introduction.}
\qquad During last years the fluxbrane compactifications with strongly warped throats were intensively used in attempts to "build a bridge" between Standard Model and Planck scales and to receive the inflaton potential providing early Universe inflation demanded by the astrophysical observations (\cite{Strominger} - \cite{Cline} and references therein). In these papers the Klebanov-Strassler \cite{Klebanov} compactification on Calabi-Yau manifold is basically used. Being inspired by this promising work we consider the predictions of the most familiar "throat-like" extremal and non-extremal $p$-brane solutions in string-based supergravity with dilaton and antisymmetric tensor \cite{Metsaev} where compactification of a number of extra dimensions is performed on a sphere $S^{n}$ \cite{Horowitz} - \cite{Aharony}. 

The isotropic coordinate $r$ along the throat plays in these solutions the role of extra coordinate in the Randall-Sundrum theory \cite{Randall}. Compactification of the throat-like solution in this direction may be realized by introduction of the co-dimension one ultraviolet brane where $Z_{2}$ identification and corresponding junction conditions are imposed \cite{Grojean}. This brane, which serves a "heavy lid" of the solution, winds over $n$-sphere and junction conditions are essentially anisotropic in different brane's subspaces. It is shown in the present paper (Sec. 2) that anisotropic junction conditions stabilize the "isotropic" co-dimension one brane at the top of the throat where extra dimensional space acquires flat asymptotic. 

The effective radion potential, which exact analytical form is received in Sec. 3 proves to be non-negative, possesses exponential and rather flat (hopefully flat enough to comply with observations \cite{Dvali}-\cite{WMAP}) behavior deep inside the throat, steep slope close to the top of the throat, and zero minimum at the position of the brane fixed by junction conditions. The minimum is separated by the barrier from the runaway decompactification to the infinite volume of extra dimensions (see Fig. \ref{AltFig1}, \ref{AltFig2}, Subsec. {\it 3-b}).

Radion field is determined conventionally as the position of the brane slowly depending on the "external" coordinates $x^{\mu}$ \cite{Gregory}. Its effective potential is calculated by the standard procedure of integrating out extra coordinates in the higher-dimensional action where elementary bulk fluxbrane solution is taken as a background. Upper limit of the integration over isotropic coordinate $r$ is given by the UV brane's position moved arbitrarily from its location fixed by junction conditions. This "moving lid" approach should not be confused 
with the "brane running" considerations \cite{Brevik} where junction conditions imposed at any brane's position serve a tool to calculate brane's effective action from RG flow equations in frames of the AdS/CFT correspondence \cite{Verlinde}. In our case situation is opposite: action of the brane is given {\it a priory} by the simplest tension 
term whereas junction conditions at the "moved" brane are violated; this results in non-trivial effective radion potential. To receive the physically meaningful radion potential the extremal magnetic monopole $p$-brane solution must be taken as a bulk background, not the dual electric one. The nonequivalence of two solutions is immediately seen when the higher-dimensional consistency conditions of the paper \cite{Leblond} (which generalize the D5 result of \cite{Kallosh}) are applied - see Appendix.

To calculate in frames of considered models the value of the electro-weak hierarchy (Sec. 4) it is necessary, as ordinary, to fix the IR end of the warped throat where visible matter is trapped. The dynamical trapping of matter on the surface was proposed in pioneer works \cite{Rubakov, Akama}. It was recently discussed however that in the strongly warped space-time the gravitational accretion of massive matter to the IR end may play the role of "trapping" \cite{Katherine}. Also the exponential growth of the massive Kaluza-Klein wave functions towards the infrared end of space-time takes place \cite{Burgess}. Falling down the throat is also true for different branes moving in fluxbrane solution as a background \cite{Giant, Branonium}. 

In any case in the present paper, which essentially develops ideas of works \cite{Altsh}, it is conventionally supposed that there is trapping or accretion of the visible massive matter at the IR end of the warped throat. To receive the numbers we shall "boldly" suppose \cite[d]{Altsh} that matter falls down to the very "bottom" of the throat, i.e. to the IR end where curvature increases to the fundamental string scale and low-energy string approximation stops to be valid (this criteria evidently does not work for models with constant curvature in the infinite throat, like e.g. in the Type $IIB$ supergravity with $AdS_{5}\times S^{5}$ asymptotic inside the throat). We shall not try to justify this questionable hypotheses but just look at its consequences. In particular it is shown that the 4-brane extremal solution in D10 Type $IIA$ supergravity may give the number close to the observed value of the electro-weak hierarchy. It must be also noted that models under consideration do not suppose any fine-tuning.

The latter 4-brane solution demands additional compactification from 4 to 3 space dimensions. This is done by considering the Euclidian "time" version of the non-extremal Schwarzshild-type modification of the extremal fluxbrane solution \cite{Horowitz}-\cite{Aharony} where the "time" torus $S^{1}$ is an additional compactified extra dimension. This solution provides mathematical tool to construct the smooth IR end of the throat at the "bolt" point (in terminology of \cite{Hawking}, see also \cite{Louko}, \cite{Aghababaie} and references therein). In this way the period of the compactified torus $S^{1}$ is determined; the stabilization of this modulus is important for calculation of values of physical quantities.

And quite naturally this generalization of the elementary fluxbrane solution may result in the acceleration of the Universe in modern epoch. The point is that this background violates Israel junction conditions at the UV brane. And if the "bolt" point (i.e. IR end) is located deep in the throat this discrepancy proves to be extremely small. In \cite[a]{Altsh}, where 6D version of the RS-model was considered, it was shown that the remedy may be the introducton of small positive cosmological constant in 4 dimensions; its estimated value proves to be close to the observed one when the Reissner-Nordstrom modification of the RS-model is used. However in the present paper where higher-dimensional models are considered we do not possess the Reissner-Nordstrom modification of the elementary fluxbrane solution with non-zero curvature in 4 dimensions. Thus in Subsec. {\it 4-d} we just make rude estimation of the absolute value of expected deviation of the radion effective potential from its zero value in the minimum when well known Schwarzshild type generalization of the extremal fluxbrane solution is used. In the seemingly most interesting case of D10 Type $IIA$ supergravity this gives the number about 60 orders above the Dark Energy density determined from the observations (see review \cite{Copeland} and references therein).
                                                                                                                                                                                                                                                                                                                                                                                                                                                                                                                                                                                                                                                                                                                                                                                                                                                                                                                                                                                                                                                                                                                                                                                                                                                                                                                                                        
Thus this paper demonstrates that well known fluxbrane solutions proves to be rich enough in its possible predictions although many questions still must be studied.

\section{Description of the model}

{\large\it 2-a. Basic action.}

\vspace{0,5cm}

Consider the following action in $D$ dimensions:

\begin{eqnarray}
\label{1}
&&S^{(D)}=M^{D-2}\int\left[R^{(D)}-\frac{1}{2}(\nabla\varphi)^2-\frac{1}{2n!}e^{\alpha\varphi}F_{n}^2\right.-  \nonumber   
\\
&&-\left.\sigma e^{\gamma\varphi}\delta^{(1)}\frac{\sqrt{-h^{(D-1)}}}{\sqrt{-g^{(D)}}}\right]\sqrt{-g^{(D)}}\,d^{D}x+\rm{GH},
\end{eqnarray}
which bulk part is an Einstein-frame truncated low-energy description of the string-based supergravity with dilaton and antisymmetric tensor \cite{Metsaev, Duff1, Aharony};
$M$ is a fundamental string mass scale ($M^{-2}\cong\alpha'$, $\alpha'$ is string length squared); $g_{AB}$, $R^{(D)}$ are metric and curvature in $D$ dimensions; $\rm{GH}$ - Gibbons-Hawking term; $F_{n}$ is $n$-form field strength; $\varphi$ - dilaton field coupled to the $n$-form with a coupling constant $\alpha$. The standard action of the co-dimension one brane is included in (\ref{1}); ${h^{(D-1)}=\det{h_{ab}}}$; $h_{ab}$ is an induced metric on the brane, $\delta^{(1)}$ - Dirac delta function fixing brane's position, mass $\sigma$ in the brane's action characterizes brane's tension $T_{\it br}$:

\begin{equation}
\label{2}
\sigma=\frac{T_{\it{br}}}{M^{D-2}}.
\end{equation}

Following \cite{Duff1} we postulate the invariance of the action (\ref{1}) under the following simultaneous scale transformation of $g_{AB}$, $\varphi$, $M$:

\begin{equation}
\label{3}
g_{AB}\to e^{2\lambda}g_{AB}, \qquad \varphi \to \varphi + \frac{2(n-1)}{\alpha}\lambda, \qquad M \to Me^{-\lambda},
\end{equation}
$\lambda=const$. This determines the dilaton-brane coupling constant $\gamma$ in (\ref{1}):

\begin{equation}
\label{4}
\gamma= -\frac{\alpha}{2(n-1)}.
\end{equation}

Because of postulated scale invariance of the action (\ref{1}) under transformations (\ref{3}) fundamental mass $M$ seemingly is of no physical meaning. But this is not the case because $M$ being a string scale determines the order of higher curvature terms not written out in the string effective action (\ref{1}). These terms do not meet the scale invariance (\ref{3}). Without loss of generality we may put $M$ equal to Planck mass, $10^{19}$ GeV.

There are two conditions of applicability of the low-energy string action (\ref{1}) (\cite{Aharony}, p. 19):
 
1) The validity of string perturbation theory demands the local string coupling, measured by $e^{\varphi}$, to be everywhere small:

\begin{equation}
\label{5}
e^{\varphi}< \,1.
\end{equation}

2) The curvature components must be small compared to the string scale:

\begin{equation}
\label{6}
|R_{...}| < \, M^{2}.
\end{equation}
For the bulk solution considered below inequality (\ref{6}) singles out the permitted region of space-time.

\vspace{0,5 cm}

{\large\it 2-b. Bulk magnetic fluxbrane solution.}

\vspace{0,5cm}

We shall consider the following elementary magnetic extremal $p$-brane solution of the dynamical equations given by the action (\ref{1}) \cite{Horowitz}-\cite{Aharony}, \cite{Grojean}:

\begin{eqnarray}
\label{7}
&&ds_{(D)}^{2}=H^{2\beta}{\widetilde{ds}}^{2}_{(p+1)}+ H^{2\xi}(dr^{2}+r^{2}d\Omega_{n}^{2}),
\\  \nonumber
&& e^{\varphi}=e^{\varphi_{\infty}}H^{\delta}, \qquad F_{n}=Qdy^{1}\wedge dy^{2}\wedge \cdots\wedge dy^{n},  \nonumber
\end{eqnarray}
where ${\widetilde{ds}}^{2}_{(p+1)}={\tilde g}_{\mu\nu}dx^{\mu}dx^{\nu}$ is the space-time interval of the manifold $M_{(p+1)}$ which in general is a product of the (3+1) dimensional observed Universe (Minkowski metric must be taken in the bulk solution (\ref{7})) and some compact Ricci-flat sub-space $K_{(p-3)}$:

\begin{equation}
\label{8}
M_{(p+1)}=M_{(3+1)}\times K_{(p-3)},
\end{equation}
$d\Omega_{n}^{2}$ in (\ref{7}) is metric of $n$-sphere of unit radius, $x^{A}=\{x^{\mu},r,y^{i}\}$, $\mu=0,1...p$; $i=1...n$; $r$ is an isotropic radial coordinate; thus $D=p+n+2$. $Q$ is charge of the magnetic monopole $n$-form field strength proportional to the volume form on $ S^{n}$; $\varphi_{\infty}$ is a value of dilaton field at $r=\infty$. Other quantities in the bulk solution (\ref{7}) are determined through dimensionalities $p, n$, coupling constant $\alpha$, and characteristic length $L$ of the solution: 

\begin{eqnarray}
\label{9}
&&\beta=-\frac{2(n-1)}{(p+n)\Delta}, \qquad \xi=\frac{2(p+1)}{(p+n)\Delta}, \qquad \delta=-\frac{2\alpha}{\Delta}, \nonumber \\
&&\Delta=\alpha^{2}+\frac{2(p+1)(n-1)}{p+n}, \qquad H=1+\left(\frac{L}{r}\right)^{n-1}.
\end{eqnarray}
Length $L$ is in turn determined through constants $Q$, $\varphi_{\infty}$:

\begin{equation}
\label{10}
L^{n-1}=\frac{\sqrt{\Delta}}{2(n-1)}\,Q\,e^{\alpha\varphi_{\infty}/2}.
\end{equation}

Metric given by (\ref{7})-(\ref{10}) describes a warped "throat" with an integrable singularity at $r=0$ (in case $\alpha \ne 0$) and $AdS_{p+2} \times S^{n}$ asymptotic with constant radius $(rH^{\xi})$ of the $n$-sphere inside the throat $(r\ll L)$ in case $\alpha = 0$. At $r>L$ space-time (\ref{7}) quickly acquires flat assymptotic.

With account of (\ref{10}) the $n$-form term in the action (\ref{1}) may be expressed only through the characteristic length of the throat $L$:

\begin{equation}
\label{11}
\frac{1}{2n!}e^{\alpha\varphi}F_{n}^2=\frac{e^{\alpha\varphi}Q^{2}}{2(H^{2\xi}r^{2})^{2n}}=\frac{2(n-1)^{2}}{H^{2\xi}r^{2}\Delta}\left[1+\left(\frac{r}{L}\right)^{n-1}\right]^{-2},
\end{equation}
thus inside the throat this term is a little bit below the value of the curvature of $n$-sphere given in the LHS of (\ref{13}).

For the bulk solution (\ref{7}) conditions (\ref{5}), (\ref{6}) of applicability of the low-energy string approximation demand:

\begin{equation}
\label{12}
\alpha > \, 0, \qquad e^{\varphi_{\infty}} < \, 1
\end{equation}
(since for $\alpha < 0$ (\ref{9}) gives $e^{\varphi} \to \infty$ at $r \to 0$), and

\begin{equation}
\label{13}
 \frac{n(n-1)}{H^{2\xi}r^{2}} < M^{2}
\end{equation}
(the positive definite curvature of $S^{n}$ is used here in the LHS). Inequality (\ref{13}) with account of expressions for $\xi$, $\Delta$ in (\ref{9}) determines the region of applicability of classical description:

\begin{equation}
\label{14}
 r > \, r_{\it min}= L\,\left[\frac{n(n-1)}{M^{2}L^{2}}\right]^{\frac{\Delta}{\alpha^{2}}}.
\end{equation}
For $\alpha=0$, when curvature is asymptotically constant inside the throat, inequality (\ref{13}) is fulfilled everywhere.

All the approach makes sense if characteristic length $L$ of the throat is essentially above the string scale $M^{-1}$:

\begin{equation}
\label{15}
ML \gg 1.
\end{equation}
Expression (\ref{10}) says that $ML$ is invariant of the scale transformation (\ref{3}).

From (\ref{15}) it also follows that $r_{\it min}$ (\ref{14}) is located somewhere deep in the throat, $r_{\it min} \ll L$. This inequality may be strengthened by the big value of exponent in the RHS of (\ref{14}); e.g. in the Type $IIA$ supergravity with 4-form in the action (\ref{1}), when $D$=10, $\alpha=1/2$, $\Delta$=4 we have $\Delta/\alpha^{2}=16$. 

In case $p=4$ in (\ref{7}) we shall use below the Schwarzshild-type non-extremal modification \cite{Horowitz}-\cite{Aharony} of the bulk solution (\ref{7}) in its Euclidian "time" version:

\begin{equation}
\label{16}
ds_{(D)}^{2}=H^{2\beta}\left({\widetilde ds}_{(3+1)}^{2}+U\left(\frac{T_{\theta}}{2\pi}\right)^{2}d\theta^{2}\right)+H^{2\xi}\left(\frac{dr^{2}}{U}+r^{2}d\Omega_{n}^{2}\right),
\end{equation}
where $\theta$ is an angle coordinate of the Euclidian "time" torus $S^{1}$ ($0<\theta<2\pi$) which period is $T_{\theta}$, and where

\begin{equation}
\label{17}
U=1-\left(\frac{r_{\it Sch}}{r}\right)^{n-1},
\end{equation}
$r_{\it Sch}$ is an arbitrary constant - analogy of Schwarzshild radius. The "bolt" point $r=r_{\it Sch}$ where $U=0$ is topologically equivalent to the pole of 2-sphere \cite{Hawking}-\cite{Aghababaie}. Space-time (\ref{16}) may possess conical singularity at this point in case "matter trapping" co-dimension two IR brane is placed there (see e.g. \cite{Aghababaie, Carroll, Rubakov2}). This will produce deficit angle $\delta_{d}$ depending on tension of the IR brane and will influence the value of period $T_{\theta}$ of the Euclidian "time" $S^{1}$ calculated from (\ref{16}), (\ref{17}):

\begin{equation}
\label{18}
T_{\theta}=\delta_{d}\,\frac{4\pi}{n-1}\,[H(r_{\it Sch})]^{\xi-\beta}r_{\it Sch}\approx \delta_{d}\,L\,\frac{4\pi}{n-1}\left(\frac{L}{r_{\it Sch}}\right)^{\left[\frac{2(n-1)}{\Delta}-1\right]},
\end{equation}
last approximate equality is valid at $r_{\it Sch}\ll L$. For $\delta_{d}=1$ (no co-dimension two brane at the "bolt") the throat of space time (\ref{16}) comes to a smooth IR end at $r=r_{\it Sch}$. As it was said in the Introduction solution (\ref{16}) provides a natural mathematical tool for compactification from 4 space dimensions of the manifold $M_{4+1}$ (in case $p=4$ in (\ref{7})-(\ref{9})) to 3 space dimensions of the observed Universe; period of the extra torus in this case is not an arbitrary modulus of the solution but is determined by expression (\ref{18}). Also metric of type (\ref{16}) may produce the extremely small Dark Energy density in modern epoch (see Sec. 4).

\vspace{0,5 cm}

{\large\it 2-c. Stabilization of the position of UV brane.}

\vspace{0,5cm}

Let us limit the volume of extra dimensions of the bulk space-time (\ref{7}) by the ultraviolet co-dimension one brane terminating the throat at some $r=r_{0}$. Thus we truncate space-time (\ref{7}) at $r=r_{0}$, paste two copies of the inner region along the cutting surface and place there brane demanding observation of $Z_{2}$-symmetry at the boundary. Brane's space-time is a product

\begin{equation}
\label{19}
M_{(p+1)} \times S^{n},
\end{equation}
thus co-dimension one brane winds over $ S^{n}$, it may be also viewed as a spherical shell of $p$-branes. Since the magnetic fluxbrane solution is considered the brane is uncharged. We suppose that brane's energy-momentum tensor is of standard isotropic form as it follows from the brane's action in (\ref{1}). Dynamical equations received by the functional variations of the action (\ref{1}) give three junction conditions at the brane: two Israel conditions for two subspaces in (\ref{19}) and jump condition for dilaton field $\varphi$. Thus we have at the brain's position $r=r_{0}$ (factor 2 in the LHS reflects the $Z_{2}$-symmetry, prime means derivation over $r$):

\begin{equation}
\label{20}
\frac{2}{H^{2\xi}} \, \beta \frac{H'}{H}=\frac{\sigma e^{\gamma\varphi}}{2(p+n)H^{\xi}},
\end{equation}

\begin{equation}
\label{21}
\frac{2}{H^{2\xi}} \, \left(\xi \frac{H'}{H}+\frac{1}{r}\right)=\frac{\sigma e^{\gamma\varphi}}{2(p+n)H^{\xi}},
\end{equation}

\begin{equation}
\label{22}
-\frac{2}{H^{2\xi}} \, \delta \frac{H'}{H}=\frac{\gamma\sigma e^{\gamma\varphi}}{H^{\xi}},
\end{equation}
where $\sigma$, $\gamma$ are given in (\ref{2}), (\ref{4}) and $\beta$, $\xi$, $\delta$, $H$ in (\ref{9}). The "scaling" value (\ref{4}) of the brane-dilaton coupling constant $\gamma$ makes condition (\ref{22}) consistent with two others. Whereas (\ref{20}), (\ref{21}) determine the brane's position $r_{0}$:

\begin{equation}
\label{23}
r_{0}=L \, \left[\frac{2(n-1)}{\Delta}-1\right]^{\frac{1}{n-1}}
\end{equation}
(in the next Section we'll see that $r_{0}$ is a point of zero minimum of the radion effective potential), and also connect $L$, $\varphi_{\infty}$ and $\sigma$:

\begin{equation}
\label{24}
L=(\sigma e^{\gamma\varphi_{\infty}})^{-1} \, 4(n-1)\left[\frac{\Delta}{2(n-1)}\right]^{\frac{1}{n-1}}.
\end{equation}

From expressions (\ref{24}) and (\ref{10}) follows the relation between magnetic monopole charge $Q$ and mass $\sigma$ where modulus $\varphi_{\infty}$ drops out:

\begin{equation}
\label{25}
Q=\frac{[4(n-1)]^{n-1} \, \sqrt{\Delta}}{\sigma^{n-1}}.
\end{equation}
This is a direct analogy of the fine-tuning of the bulk cosmological constant and brane's tension demanded in the Randall-Sundrum model \cite{Randall}. However here the bulk magnetic fluxbrane charge $Q$ is not an input parameter in the action (\ref{1}) but a free constant of the bulk solution of the dynamical equations. Hence relation (\ref{25}) is by no means a fine-tuning but just a constraint determining magnetic charge $Q$ through the brane's tension.

From (\ref{24}) and (\ref{4}) it also follows the important expression for the basic dimensionless parameter $ML$ of the considered solutions:

\begin{equation}
\label{26}
ML=g \, 4(n-1)\left[\frac{\Delta}{2(n-1)}\right]^{\frac{1}{n-1}},
\end{equation}
where

\begin{equation}
\label{27}
g=\frac{M}{\sigma} \, e^{\frac{\alpha}{2(n-1)}\varphi_{\infty}}.
\end{equation}
We'll see that dimensionless parameter $g$, which is an invariant of scale transformation (\ref{3}), essentially determines many physical predictions of the theory. It will be argued below that it is natural to take $g=1$ ({\it Hyp. 3} in Subsec. {\it 4-b}). From expression (\ref{26}) it is seen that for the sensible values of dimensionality $n$ ($n\ge 4$) strong inequality (\ref{15}) for $ML$ is fulfilled even for $g=1$. For example in the Type $IIA$ supergravity with 4-form field strength (when $p=4$, $n=4$, $\alpha=1/2$) relation (\ref{26}) gives $ML=10,48 \cdot g$.

The choice of modulus $\varphi_{\infty}$ (or equivalently the choice of $g$ and $L$, see (\ref{27}), (\ref{26})) fix actually the volume of extra space in the fluxbrane solution (\ref{7}) since according to (\ref{23}) characteristic length of the throat $L$ determines the position $r_{0}$ where the "lid" brane is stabilized by the dynamical equations. In calculating the radion effective potential we may fix the brane's position $r_{0}$ by the choice of some basic value of $\varphi_{\infty}$ and then change arbitrarily the modulus $\varphi_{\infty}$, supposing its slow dependence on space-time coordinates $x^{\mu}$. Or vice versa: we may fix the background value of $\varphi_{\infty}$ (i.e. fix $g$, $L$) and arbitrarily move UV brane from its given stable position $r=r_{0}$ (\ref{23}). This "moving lid" approach proves to be more transparent and convenient, we shall use it in the next section.

It must be noted that for the non-extremal metric (\ref{16}) Israel junction condition (\ref{20}) splits into two for $M_{3+1}$ and for the torus $ S^{1}$:

\begin{equation}
\label{28}
\frac{2U}{H^{2\xi}} \, \beta \frac{H'}{H}=\frac{\sigma e^{\gamma\varphi}\sqrt{U}}{2(p+n)H^{\xi}},
\end{equation}

\begin{equation}
\label{29}
\frac{2U}{H^{2\xi}} \, \left(\beta \frac{H'}{H}+\frac{U'}{2U}\right)=\frac{\sigma e^{\gamma\varphi}\sqrt{U}}{2(p+n)H^{\xi}},
\end{equation}
here $r=r_{0}$ must be taken, $U(r)$ is given in (\ref{17}), prime is derivative over $r$. Evident discrepancy of (\ref{28}) and (\ref{29}) is quite small in case $r_{\it Sch}\ll r_{0}$. We'll see below (Subsec. {\it 4-d}) that this discrepancy may show itself in appearance of the extremely small deviation of the radion effective potential from its zero value in the minimum.

To finalize the description of the model we specify again that when "lid" brane is stabilized in the minimum of the radion potential (this supposedly happens after early stages of inflation and reheating) the UV end of the throat is located approximately at the top of the throat at $r=r_{0}$ (\ref{23}), whereas the IR end of the throat, $r=r_{\it IR}$ (where supposedly Standard Model resides), needs to be fixed by hand - by the introduction of the negative tension "anisotropic" IR brane or by the "bolt" tool as in space-time (\ref{16}). In any case $r_{\it IR}$ must exceed the minimal permitted value (\ref{14}) of the isotropic coordinate $r$:

\begin{equation}
\label{30}
r_{\it min}<r_{\it IR}\le r \le r_{0}.
\end{equation}

Like in all RS-type models the choice of $r_{\it IR}$ determines the calculated value of the electro-weak mass scale hierarchy (see Sec. 4).

\section{"Moving lid" approach gives the promising \\
form of the radion effective potential}

{\large\it 3-a. Effective Brans-Dicke action in $(p+1)$ dimensions.}

\vspace{0,5cm}

Effective action in lower dimensions is conventionally received by integrating out extra coordinates in a higher dimensional action. To calculate the effective action $ S^{(p+1)}$ we shall use in (\ref{1}) the bulk solution (\ref{7}) but move the UV "lid" brane (and hence change the upper limit of the integration over isotropic coordinate $r$ in (\ref{1})) from $r=r_{0}$ (\ref{23}) fixed by junction conditions (\ref{20})-(\ref{22}) to the arbitrary position $\rho (x)$, slowly depending on coordinates $x^{\mu}$:

\begin{equation}
\label{31}
r_{0} \to \rho (x),
\end{equation}
$\rho (x)$ is called radion field \cite{Gregory}. Its gradient terms contribute to the brane's induced metric:

\begin{equation}
\label{32}
h_{ab}=g_{ab}+\rho,_{a}\rho,_{b}g_{rr},
\end{equation}
where $x^{a}=\{x^{\mu}, y^{i}\}$ and $g_{ab}$, $g_{rr}$ are corresponding components of the bulk metric (\ref{7}). Then, with account that $\rho(x)$ does not depend on $y^{i}$ and depends on $x^{\mu}$ slowly as compared to the scales of the bulk solution, the brane's Lagrangian in (\ref{1}) takes the form:

\begin{eqnarray}
\label{33}
&&L_{\it br}=-M^{D-2}\sigma e^{\gamma\varphi}\delta(r-\rho)\,\frac{\sqrt{-h^{(D-1)}}}{\sqrt{-g^{(D)}}}=  \nonumber
\\ \nonumber
&&=-\frac{M^{D-2}\sigma e^{\gamma\varphi}\,\delta(r-\rho)}{H^{\xi}\sqrt{-{\tilde g}^{(p+1)}}}\sqrt{-\det{\left[{\tilde g}_{\mu\nu}+H^{2(\xi-\beta)}\rho,_{\mu}\rho,_{\nu}\right]}}
\\
&&\approx -\frac{M^{D-2}\sigma e^{\gamma\varphi}\,\delta(r-\rho)}{H^{\xi}}\left[1+\frac{1}{2}H^{2(\xi-\beta)}{\tilde g}^{\mu\nu}\rho,_{\mu}\rho,_{\nu}\right],
\end{eqnarray}
${\tilde g}^{(p+1)}=\det{{\tilde g}_{\mu\nu}}$.

In calculating effective action we shall substitute $r_{\it IR}\to r=0$ in the lower limit of the integration over $r$. This will not change $S^{(p+1)}$ essentially since it is supposed that $r_{\it IR}\ll L$ and all integrals are convergent in the singular point $r=0$. We also postulate that $Z_{2}$-symmetry at the "moved" brane is preserved, i.e. bulk integration must be fulfilled over two pasted copies of the solution; as expected this procedure gives zero value of the radion effective potential when $\rho=r_{0}$. Thus symbolically the following Brans-Dicke type effective action $S^{(p+1)}$ depending on general metric ${\tilde g}_{\mu\nu}(x)$ of the manifold $M_{(p+1)}$ and radion field $\rho (x)$ is received from the action (\ref{1}):

\begin{eqnarray}
\label{34}
&&S^{(p+1)}=2\int_{0}^{\rho}L_{\it bulk}+\int L_{\it br}=
\\
&&=\int\left[\Phi(\rho){\widetilde R}^{(p+1)}-\frac{1}{2}\omega (\rho){\tilde g}^{\mu\nu}\rho,_{\mu}\rho,_{\nu}-{\widetilde V}(\rho)\right]\sqrt{-{\tilde g}^{(p+1)}}d^{p+1}x, \nonumber
\end{eqnarray}
where $L_{\it bulk}$ sums up all non-brane terms in (\ref{1}) including the Gibbons-Hawking term, $L_{\it br}$ is given in (\ref{33}); ${\widetilde R}^{(p+1)}$ is scalar curvature of the $(p+1)$ dimensional space-time described by metric ${\tilde g}_{\mu\nu}$. Brans-Dicke field $\Phi (\rho)$, kinetic term function $\omega (\rho)$ and auxiliary radion potential ${\widetilde V}(\rho)$ are calculated when bulk metric (\ref{7}) is used in (\ref{1}) where it is taken ${\tilde g}_{\mu\nu}=\eta_{\mu\nu}$ - Minkowski metric in $(p+1)$ dimensions; also formulae (\ref{11}), (\ref{24}) permitting to express the $n$-form and brane terms of the action (\ref{1}) through the characterictic length $L$ of the bulk solution (\ref{7}) were taken into account. Simple calculations finally give:

For the Brans-Dicke field in the action (\ref{34}):

\begin{eqnarray}
\label{35}
&&\Phi(\rho)=2M^{D-2}\Omega_{n}\int_{0}^{\rho} H^{2(\xi - \beta)}r^{n}\,dr \, = \, 2M^{p-1}(ML)^{n+1}\Omega_{n}f\left(\frac{\rho}{L}\right), \nonumber
\\
&&f(y)=\int_{0}^{y}\left(1+\frac{1}{\zeta^{n-1}}\right)^{\frac{4}{\Delta}}\zeta^{n}\,d\zeta.
\end{eqnarray}

For the kinetic term function:

\begin{eqnarray}
\label{36}
&&\omega(\rho)=M^{D-2}\Omega_{n}\int \sigma e^{\gamma\varphi}\delta(r-\rho)H^{(3\xi - 2\beta)}r^{n}\,dr= \nonumber \\
\\  
&&=M^{p+1}(ML)^{n-1}\Omega_{n}\,4(n-1)\left[\frac{\Delta}{2(n-1)}\right]^{\frac{1}{n-1}}\left(\frac{\rho}{L}\right)^{n}\left[1+\left(\frac{L}{\rho}\right)^{n-1}\right]^{\kappa},  \nonumber
\\  \nonumber
&&\kappa=\frac{1}{n-1}+\frac{4}{\Delta}.
\end{eqnarray}

And for potential in (\ref{34}):

\begin{eqnarray}
\label{37}
&&{\widetilde V}(\rho)=M^{D-2}\Omega_{n}\left\{-2\int_{0}^{\rho}\left[\frac{2n(n-1)}{H^{2\xi}r^{2}}-\frac{Q^{2}e^{\alpha\varphi}}{H^{2n\xi}r^{2n}}\right]H^{2\xi}r^{n}\,dr\right.+ \nonumber
\\
&&+\left.\int \sigma e^{\gamma\varphi}\delta(r-\rho)H^{\xi}r^{n}\,dr\right\}=4M^{p+1}(ML)^{n-1}\Omega_{n}F\left(\frac{\rho}{L}\right).
\end{eqnarray}
In (\ref{35})-(\ref{37}) $\Omega_{n}$ is volume of $n$-sphere of unit radius. Function $F(y)$ in (\ref{37}) is given by the formula:

\begin{eqnarray}
\label{38}
&&F(y)=y^{n-1}\left[(n-1)\left(\frac{1+y^{n-1}}{q}\right)^{\frac{1}{n-1}}+\left(\frac{1+y^{n-1}}{q}\right)^{-1}-n\right], \nonumber
\\
&&y=\frac{\rho}{L}, \qquad  q=\frac{2(n-1)}{\Delta}>1.
\end{eqnarray}

According to (\ref{23}) position of the brane fixed by junction conditions (\ref{20})-(\ref{22}) corresponds to the value of $y$:

\begin{equation}
\label{39}
y_{0}=\frac{r_{0}}{L}=(q-1)^{\frac{1}{n-1}},
\end{equation}
$q$ is defined in (\ref{38}). It is easy to see that $F(y)$ possesses minimum at $y=y_{0}$ and $F(y_{0})=0$. The same is true for potential ${\tilde V}(\rho)$ (\ref{37}) at $\rho=r_{0}$.

{\it Note 1.} Although Gibbons-Hawking term is a full divergence and we consider compact extra space it would be mistake to discard GH term in (\ref{1}) when radion effective potential is calculated. GH contribution to the radion potential really vanish at the solution of dynamical equations, i.e. at $\rho=r_{0}$, but it is by no means equal to zero when upper limit of integration in (\ref{34}) is changed from $r_{0}$ to arbitrary value $\rho$.

{\it Note 2.} It is impossible to calculate from the action (\ref{1}) the physically meaningful radion effective potential in case dual electric $(p+2)$-form $F_{(p+2)}$ is used in (\ref{1}), i.e. when electric $p$-brane solution is taken as a background. Although electric $p$-brane extremal solution is given by the same formulae (\ref{7})-(\ref{10}), (\ref{23})-(\ref{25}) as a magnetic one, the values of action (\ref{1}), $ S_{m}$ and $ S_{e}$, calculated at the magnetic and electric fluxbrane solutions correspondingly drastically differ. General consistency conditions \cite{Leblond} say that $ S_{m}$ must vanish but they are not applicable to $ S_{e}$ (see Appendix).

According to (\ref{37}), (\ref{38}) ${\widetilde V}(\rho)\to 0$ at $\rho\to 0$ and ${\widetilde V}(\rho)\to \infty$ at $\rho\to \infty$. However this behavior is of no physical interest since similar behavior possesses the Brans-Dicke field (\ref{35}). To get the physically meaningful radion effective potential the low dimension Brans-Dicke effective action (\ref{34}) must be written in the Einstein-frame metric and radion field must be transformed in a way providing the canonical form of its kinetic term.

\vspace{0,5 cm}

{\large\it 3-b. Einstein-frame effective action in $(p+1)$ dimensions.}

\vspace{0,5cm}

Let us rescale metric ${\tilde g}_{\mu\nu}$ in the Brans-Dicke action in the RHS of (\ref{34}) to the Einstein-frame metric $g_{\mu\nu}$:

\begin{equation}
\label{40}
{\tilde g}_{\mu\nu}=\frac{M_{0}^{2}}{\left[\Phi(\rho)\right]^{\frac{2}{p-1}}}\,g_{\mu\nu},
\end{equation}
where Planck mass in $(p+1)$ dimensions was introduced:

\begin{equation}
\label{41}
M_{0}\equiv M_{{\rm Pl}\,(p+1)}.
\end{equation}

Effective action (\ref{34}) being expressed as a functional of the Einstein-frame metric $g_{\mu\nu}$ and canonical radion field $\psi$ (defined below) takes the standard form:

\begin{equation}
\label{42}
S^{(p+1)}=\int \left[M_{0}^{p-1}R^{(p+1)}-(1/2)M_{0}^{p-1}(\nabla\psi)^{2}-\mu^{p+1}V(\psi)\right]\sqrt{-g^{(p+1)}}\,d^{p+1}x,
\end{equation}
$\mu$ is a calculable constant of dimensionality of mass - the characteristic of the radion potential, $V(\psi)$ is taken dimensionless for convenience. Also dimensionless (normalized to Planck mass (\ref{41})) canonical radion field $\psi(\rho)$ is introduced in (\ref{42}):

\begin{equation}
\label{43}
\psi(\rho)=\frac{1}{L}\int_{r_{0}}^{\rho}\epsilon_{(p)}(\rho)\,d\rho=\int_{y_{0}}^{y}\epsilon_{(p)}(y)\,dy, \qquad y=\frac{\rho}{L},
\end{equation}
here the point (\ref{23}) of stable extremum of the radion effective potential is chosen at $\psi=0$; $y_{0}$ see in (\ref{39}); $\epsilon_{(p)}$ is expressed through functions $\Phi(\rho)$, $\omega(\rho)$ given in (\ref{35}), (\ref{36}) (subscript $(p)$ is introduced for later usage):

\begin{eqnarray}
\label{44}
&&\epsilon_{(p)}^{2}=L^{2}\left[\frac{\omega(\rho)}{\Phi(\rho)}+\frac{2p}{p-1}\left(\frac{1}{\Phi}\frac{d\Phi}{d\rho}\right)^{2}\right]=
\\ \nonumber
&&=\frac{2(n-1)y^{n}}{f(y)}\left(\frac{\Delta}{2(n-1)}\right)^{\frac{1}{n-1}}\left(1+\frac{1}{y^{n-1}}\right)^{\kappa}+\frac{2py^{2n}}{(p-1)f^{2}(y)}\left(1+\frac{1}{y^{n-1}}\right)^{\frac{8}{\Delta}}, \nonumber
\end{eqnarray}
where function $f(y)$ and constants $\kappa$, $\Delta$ are given in (\ref{35}), (\ref{36}), (\ref{9}) correspondingly.

It is seen from (\ref{44}) that in the $\rho \ll L$ ($y \ll 1$) limit $\epsilon(y) \sim y^{-1}$ and in the $\rho \gg L$ ($y \gg 1$) limit we have $\epsilon \sim y^{-1/2}$. Hence it follows from (\ref{43}) that in these two limits:

\begin{eqnarray}
\label{45}
&&\psi=c_{(p)}\ln y, \qquad 0<y=\frac{\rho}{L}\ll 1,
\\
&&c_{(p)}=\left\{2(n+1-2q)\left[(n-1)q^{-\frac{1}{n-1}}+\frac{p}{p-1}(n+1-2q)\right]\right\}^{1/2}, \nonumber
\end{eqnarray}
and

\begin{equation}
\label{46}
\psi=[8(n^{2}-1)q^{-\frac{1}{n-1}}]^{1/2}\,y^{1/2}, \qquad  1\ll y=\frac{\rho}{L}<\infty,
\end{equation}
$q$ see in (\ref{38}).

Radion potential $\mu^{p+1}V(\psi)$ in (\ref{42}) is expressed through the auxiliary potential ${\widetilde V}(\rho)$ (\ref{37}) and Brans-Dicke field $\Phi (\rho)$ (\ref{35}):

\begin{equation}
\label{47}
\mu^{p+1}V(\psi)=M_{0}^{p+1}{\widetilde V}(\rho)[\Phi(\rho)]^{-\frac{p+1}{p-1}},
\end{equation}
where dependence $\rho(\psi)$ is deduced from (\ref{43}). Finally:

\begin{eqnarray}
\label{48}
&&\mu^{p+1}V(\psi)=M_{0}^{p+1}\left[\frac{2^{p-3}}{(ML)^{2(p+n)}\Omega_{n}^{2}}\right]^\frac{1}{p-1}K(y(\psi)),
\\ \nonumber
&&V(\psi)\equiv K(y(\psi))=F(y)\left[f(y)\right]^{-\frac{p+1}{p-1}}, \nonumber
\end{eqnarray}
$F(y)$, $f(y)$ are determined in (\ref{38}), (\ref{35}) and we remind that $y(\psi)=\rho(\psi) / L$ must be found from (\ref{43}), (\ref{44}). 

Characteristic mass $\mu$ defined by (\ref{48}) is suppressed as compared to the Planck mass (\ref{41}) because of the strong inequality (\ref{15}). Thus radion potential (\ref{48}) may meet demand of applicability of the low-energy approximation (it must be below Planck density) in rather wide region of the radion field $\psi$. We'll come back to this point in Subsec. {\it 4-c}.

It is most essential that dimensionless potential $V(\psi)$ in (\ref{48}) depends only on dimensionalities and coupling constant $\alpha$ in (\ref{1}), i.e. it depends on the choice of the theory, not on the arbitrary constant $(ML)$ (\ref{26}) (or $g$ (\ref{27})) of the fluxbrane solution (\ref{7}).

\begin{figure}
\vspace{-3cm}
\begin{center}
\includegraphics[width=12cm]{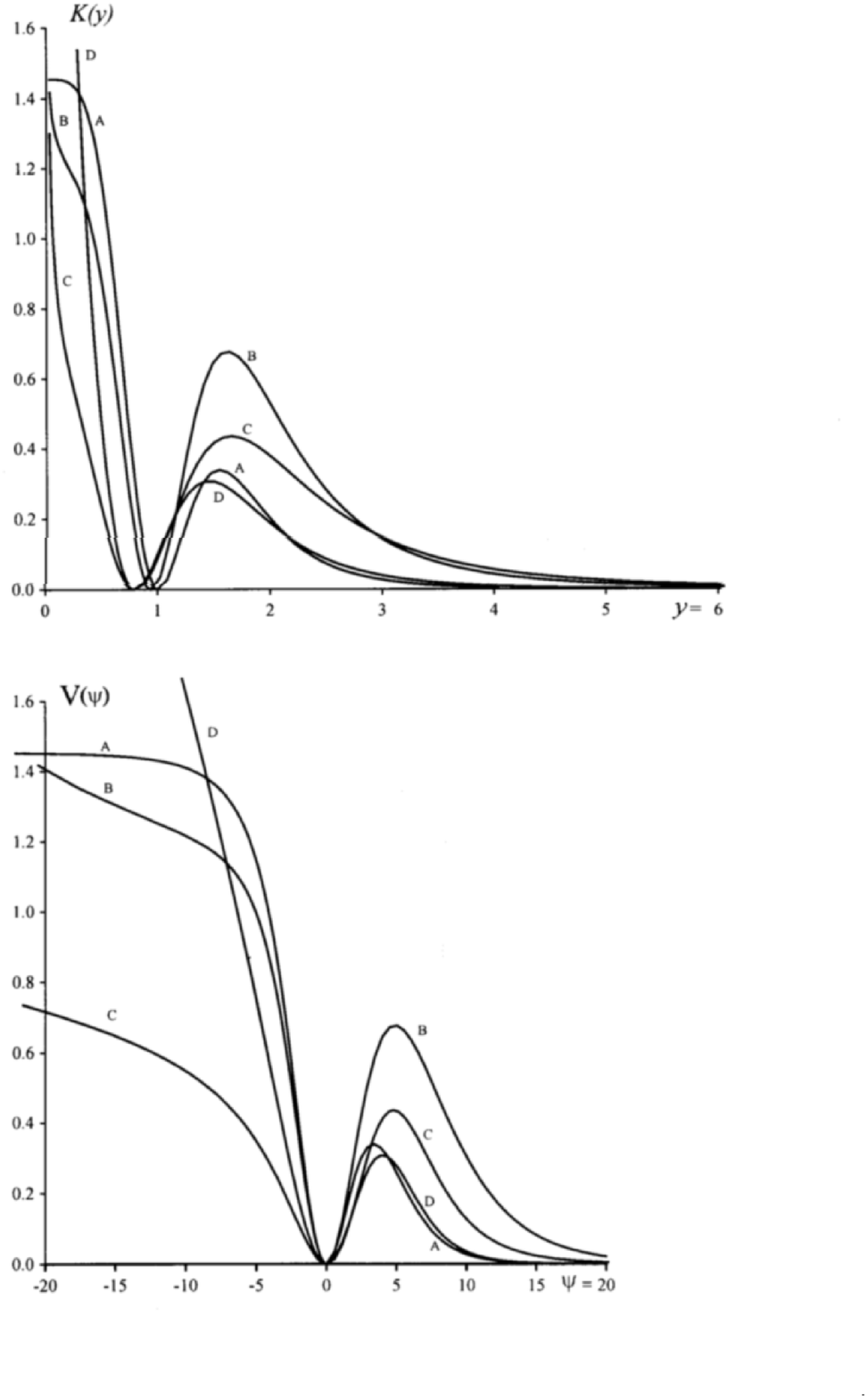}
\end{center}
\caption{\footnotesize {Curves {\it A}, {\it B}, {\it C}  present analytical dependences
(\ref{48}) of the dimensionless effective radion potential $V(\psi)$ and
auxiliary function $K(y)$ for three theories listed in (\ref{49}). Curve
{\it D} shows dependences (\ref{57}) of the dimensionless effective radion
potential $V_{(3)}(\psi)$ and auxiliary function $K_{(3)}(y)$ for the Type
$IIA$ supergravity (theory {\it C}  in (\ref{49})) compactified to (3+1)
dimensions.}}
\label{AltFig1}
\end{figure}

Fig. \ref{AltFig1} shows dependences of $V(\psi)$ and $K(y)$ in (\ref{48}) for three theories which Bose-sector is described (partly) by the action (\ref{1}):

\begin{eqnarray}
\label{49}
&&{\it A)} \hspace{0,5cm} D=10, \quad p=3, \quad n=5, \quad \alpha=0; \nonumber  \\
&&{\it B)} \hspace{0,5cm} D=11, \quad p=4, \quad n=5, \quad \alpha=(3\sqrt{2})^{-1};  \\
&&{\it C)} \hspace{0,5cm} D=10, \quad p=4, \quad n=4, \quad \alpha=1/2. \nonumber
\end{eqnarray}
(Curve {\it "D"} at Fig. \ref{AltFig1} relates to theory {\it C} additionally compactified from 4 to 3 space dimensions - see next subsection). Theories {\it A}, {\it C} are subsectors of the Type $IIB$ and $IIA$ supergravities. Theory {\it B} in (\ref{49}) with the intermediate value of coupling constant $\alpha$ is included for illustrative purposes; it is the version of D11 string-based supergravity considered in \cite[d]{Altsh}.

\begin{figure}
\begin{center}
\includegraphics[width=15cm]{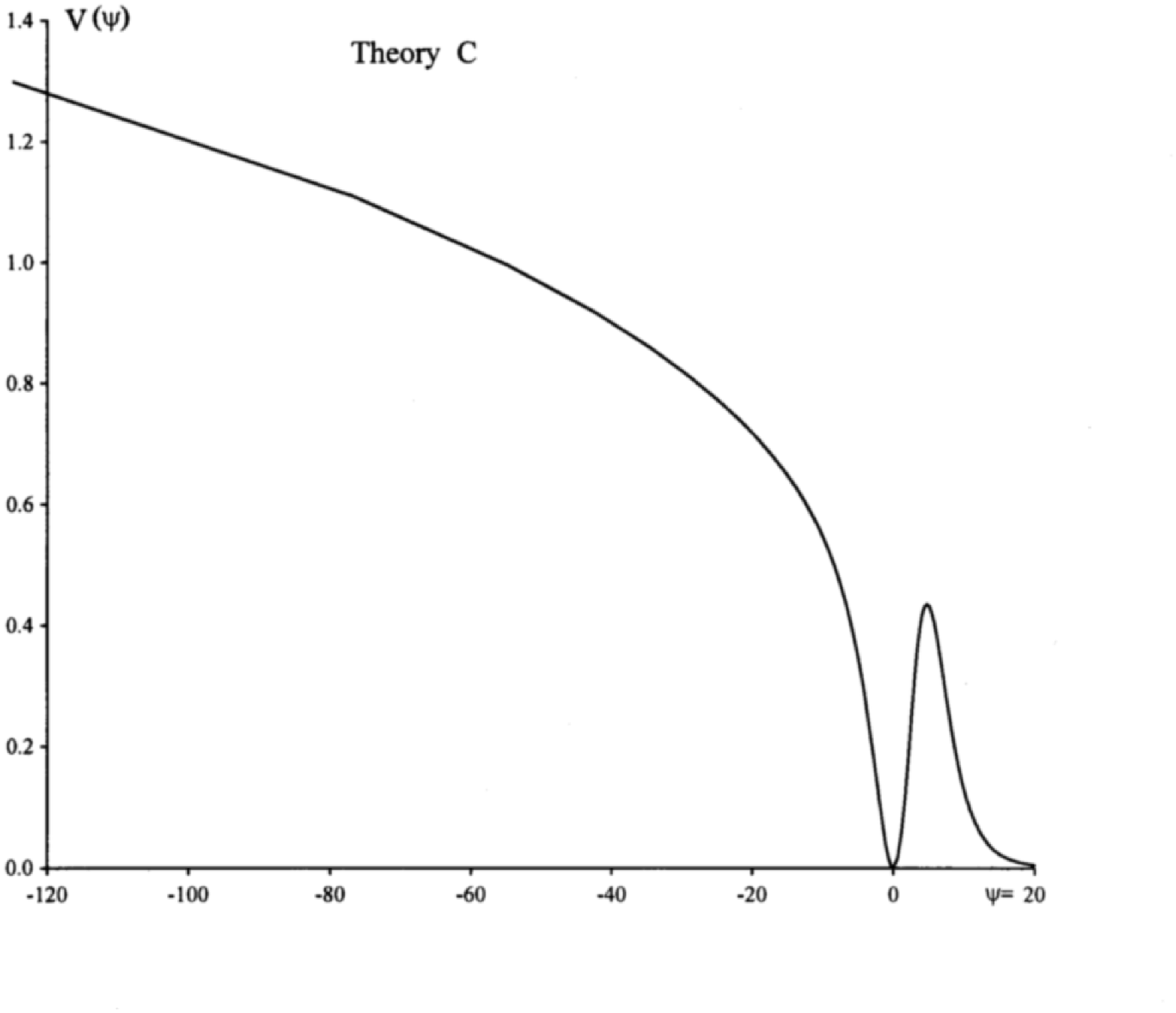}
\end{center}
\caption{\footnotesize {Behavior of the dimensionless effective radion potential $V(\psi)$
(\ref{48}) deep in the throat of the 4-brane solution (\ref{7}) for the
Type $IIA$ supergravity (theory {\it C} in (\ref{49})).}}
\label{AltFig2}
\end{figure}

For the supposedly most interesting theory {\it C} Fig. \ref{AltFig2} shows behavior of $V(\psi)$ (\ref{48}) deep in the throat up to $\psi=-120$ (according to the asymptotic (\ref{45}) applied to the theory {\it C} this corresponds to $y=\rho/L=10^{-11}$). It must be noted that scale of $V$ at Fig. \ref{AltFig1} is 25 times as much as that of abscissa $\psi$, and 100 times at Fig. \ref{AltFig2}. Thus curves at the figures are actually much more flat then it is seen at the pictures.

As expected potential (\ref{48}) possesses zero minimum at $\psi=0$ where junction conditions (\ref{20})-(\ref{22}) are valid. To the right of this point $V(\psi)$ have barrier which protects the background state $\psi=0$ where supposedly our Universe "lives in" from the runaway decompactification. This situation is typical for all theories with compactified extra dimensions \cite{Giddings2}. It is not without interest to study in frames of considered models to what extent this "protection" is reliable. But we'll leave this work for future. 

At $\psi<0$ potential (\ref{48}) uniformly increases with decrease of $\psi$, it comes to exponential asymptotic if $\alpha \ne 0$ in (\ref{1}) and to constant in case $\alpha=0$. 

Asymptotic behavior of the dimensionless radion potential $V(\psi)$ in the limits $\psi \ll -1$ and $\psi \gg 1$ follow from (\ref{48}) with account of expressions (\ref{35}), (\ref{38}) for $f(y)$, $F(y)$ and asymptotics (\ref{45}), (\ref{46}).

Thus at $\psi \ll -1$:

\begin{eqnarray}
\label{50}
&&V(\psi)=v_{(p)}e^{-k\psi}, \qquad  k=\frac{1}{c_{(p)}} \, \frac{2(p+n)\alpha^{2}}{(p-1)\Delta},
\\  \nonumber
&&v_{(p)}=(n+1-2q)^{\frac{p+1}{p-1}}\,[(n-1)q^{-\frac{1}{n-1}}+q-n], \nonumber
\end{eqnarray}
($c_{(p)}$, $q$ are given in (\ref{45}) and (\ref{38})).

And at $\psi \gg 1$:

\begin{equation}
\label{51}
V(\psi)\sim \psi^{-\frac{2(p+2n+1)}{p-1}}.
\end{equation}

Small values of the exponent $k$ in (\ref{50}) for three theories (\ref{49})

\begin{equation}
\label{52}
k_{{A\it }}=0,  \qquad  k_{{\it B}}=\frac{1}{78},  \qquad  k_{{\it C}}=\frac{1}{14}
\end{equation}
also show that radion potential is essentially flat inside the throat of the elementary magnetic fluxbrane solution (\ref{7}).

For the Type $IIB$ supergravity (theory {\it A} in (\ref{49})) expressions (\ref{48}), (\ref{50}) and (\ref{26}), (\ref{27}) give following constant asymptotic of the radion effective potential in the throat:

\begin{equation}
\label{53}
\mu_{\it A}^{4}V_{{\it A}(-)}=\left[2,6 \cdot 10^{-3}M_{\rm Pl}\left(\frac{\sigma}{M}\right)^{2}\right]^{4},
\end{equation}
where we put $M_{0}=M_{\rm Pl}$ in (\ref{48}). It is seen that for $\sigma=M$ (when it is supposed that brane's tension in (\ref{1}) is of fundamental scale) this model is consistent, i.e. $\mu_{\it A}^{4}V_{{\it A}(-)} \ll M_{\rm Pl}^{4}$.

Contrary to the Type $IIB$ supergravity where (\ref{42}) gives effective action in 4 dimensions, theories {\it B}, {\it C} in (\ref{49}) demand further compactification from 4 to 3 space dimensions. Only after that they may be applied to reality, to description of the early inflationary Universe in particular. We'll see that compactfication to 3 dimensions essentially changes behavior of the effective radion potential inside the throat.

\vspace{0,5 cm}
{\large\it 3-c. Reduction to 3 space dimensions.}

\vspace{0,5cm}

If $p>3$ in the elementary fluxbrane solution (\ref{7}) then instead of the Brans-Dicke effective action (\ref{34}) in $(p+1)$ dimensions we must consider the effective action in (3+1) dimensions when compact manifold $K_{(p-3)}$ in (\ref{8}) is also integrated out in the action (\ref{1}):

\begin{equation}
\label{54}
S^{(3+1)}=\int\left[\Phi_{(3)}(\rho){\widetilde R}^{(3+1)}-\omega_{(3)}(\rho){\tilde g}^{\mu\nu}\rho,_{\mu}\rho,_{\nu}-{\widetilde V}_{(3)}(\rho)\right]\sqrt{-{\tilde g}^{(3+1)}}\,d^{3+1}x,
\end{equation}
here ${\tilde g}_{\mu\nu}$ is metric in 4 dimensions; $\mu, \nu = \{0, 1, 2, 3\}$; subscript $(3)$ outlines that corresponding functions in (\ref{54}) differ from those in (\ref{34}). The difference at this stage is not so great however: to receive expressions for $\Phi_{(3)}$, $\omega_{(3)}$, ${\widetilde V}_{(3)}$ it is sufficient to insert the factor $T_{(p-3)}$ (volume of compact extra space $K_{(p-3)}$) into the RHS of (\ref{35})-(\ref{37}). Important new reality comes out at the stage of scale transformation of metric ${\tilde g}_{\mu\nu}$ in (\ref{54}) to the Einstein frame metric $g_{\mu\nu}$ in 4 dimensions:

\begin{equation}
\label{55}
{\tilde g}_{\mu\nu}=\frac{M_{\rm Pl}^{2}}{\Phi_{(3)}(\rho)}\,g_{\mu\nu},
\end{equation}
where $M_{\rm Pl}=10^{19}GeV$. Contrary to the scale transformation (\ref{40}) where power of $\Phi(\rho)$ includes the dependence on dimensionality $p$ which in turn enters the elementary fluxbrane solution (\ref{7}), now in (\ref{55}) power of $\Phi_{(3)}(\rho)$ is taken for $p=3$ in (\ref{40}) whereas solution (\ref{7}) is determined for $p>3$. This is the source of main difference from the situation described in previous subsection.

Under scale transformation (\ref{55}) effective Brans-Dicke action (\ref{54}) takes the standard Einstein-frame form:

\begin{equation}
\label{56}
S^{(3+1)}=\int\left[M_{\rm Pl}^{2}R^{(3+1)}-(1/2)M_{\rm Pl}^{2}(\nabla \psi)^{2}-\mu_{(3)}^{4}V_{(3)}(\psi)\right]\sqrt{-g^{(3+1)}}\,d^{3+1}x.
\end{equation}

Canonical radion field $\psi(x)$ is again taken dimensionless (in Planck units) and is determined by (\ref{43}) where $\epsilon_{(3)}$ is given by (\ref{44}) for $p=3$. Asymptotic of $\psi(\rho)$ inside the throat is again given by (\ref{45}) where coefficient $c_{(3)}$ must be used. However the effective radion potential in (\ref{56}) now essentially differs from that in (\ref{47}), (\ref{48}):

\begin{eqnarray}
\label{57}
&&\mu_{(3)}^{4}V_{(3)}(\psi)=M_{\rm Pl}^{4}\,\frac{{\widetilde V}_{(3)}(\rho)}{[\Phi_{(3)}(\rho)]^{2}}=\frac{M_{\rm Pl}^{4}}{(M^{p-3}T_{(p-3)})(ML)^{n+3}\Omega_{n}}\,K_{(3)}(y(\psi)), \nonumber
\\
&&V_{(3)}(\psi) \equiv K_{(3)}(y(\psi))=\frac{F(y)}{f^{2}(y)},
\end{eqnarray}
where $y=\rho/L$, $T_{(p-3)}$ is volume of extra space $K_{(p-3)}$ (\ref{8}), functions $F(y)$, $f(y)$ are given in (\ref{38}) and (\ref{35}), and dependence $y(\psi)$ is found from (\ref{43}), (\ref{44}) where $\epsilon_{(p)}\to \epsilon_{(3)}$ (we note that extra volume $T_{(p-3)}$ drops out in the expression (\ref{44}) for $\epsilon_{(3)}$). 

The qualitative behavior of the dimensionless potential $V_{(3)}(\psi)$ (\ref{57}) is the same like of potential $V(\psi)$ (\ref{48}); this is seen from the curve {\it D} at Fig.\ref{AltFig1}. However at $\psi<0$ potential is now more steep (cf. curves {\it C} and {\it D}). Expressions (\ref{35}), (\ref{38}) for $f(y)$, $F(y)$ and asymptotic (\ref{45}) for $\psi(\rho)$ (where $p=3$ is taken) give asymptotic of $V_{(3)}(\psi)$ deep in the throat:

\begin{eqnarray}
\label{58}
&&V_{(3)}(\psi)=v_{(3)}\,e^{-k_{(3)}\psi}, \qquad \psi \ll -1,
\\ \nonumber
&&k_{(3)}=\frac{n+3-4q}{c_{(3)}}=\frac{1}{c_{(3)}}\,\frac{(n+3)(p+n)\alpha^{2}+2(n-1)^{2}(p-3)}{(p+n)\Delta}, \nonumber
\\ \nonumber
&&v_{(3)}=(n+1-2q)^{2}\left[(n-1)q^{-\frac{1}{n-1}}+q-n \right],  \nonumber
\end{eqnarray}
$c_{(3)}$ is calculated from (\ref{45}) for $p=3$, definition of $q$ see in (\ref{38}). It is seen that asymptotic (\ref{58}) is more steep than that of (\ref{50}) ($k_{(3)}>k$). For example for the Type $IIA$ supergravity (theory {\it C} in (\ref{49})) compactified to (3+1) dimensions expression (\ref{58}) gives:

\begin{equation}
\label{59}
V_{{\it D}(3)}(\psi)=0,44 \cdot e^{-0,21\cdot\psi},  \qquad  \psi \ll -1, 
\end{equation}
thus exponent $k_{{\it D}(3)}=0,21$ calculated from (\ref{58}) is about 3 times above the exponent $k_{{\it C}}$ in (\ref{52}). Expression (\ref{59}) gives the asymptotic of curve {\it D} at Fig.\ref{AltFig1} far at the negative $\psi$. 

Now we can look shortly at the possibility to apply these results to the description of inflation in the early Universe \cite{Guth}.

Radion field introduced above hopefully may serve as an inflaton field. We may suppose that initially "lid" brane was located somewhere deep in the throat ($\rho_{\it in} \ll L$ or $\psi_{\it in} \ll -1$) and after that, obeying the dynamics determined by the action (\ref{56}), it rolls down the exponential asymptotic (\ref{58}) (specifically (\ref{59})) of the radion potential (\ref{57}) to the steep slope leading to stable brane's position (\ref{23}) ($\psi=0$) at the top of the throat.

The following questions must be answered: Does radion potential $V_{(3)}(\psi)$ (\ref{57}) meet the necessary flatness and slow roll conditions? Can this scenario provide the number of $e$-foldings $N_{e}$ during inflation demanded by the astrophysical observations ($N_{e} \approx 80-100$)? \cite{Dvali}-\cite{WMAP} For the exponential potential like in (\ref{58}) flatness and slow roll conditions demand $k_{(3)}^{2} \ll 1$ which is more or less satisfied for $k_{(3)}=0,21$ like e.g. in (\ref{59}). The number of $e$-foldings for exponential potential is given by simple formula \cite{Dvali, Mukhanov} (prime means derivative over $\psi$ which, we remind, is dimensionless - in Planck units):

\begin{equation}
\label{60}
N_{e}=\int_{\psi_{\it in}}^{\psi_{\it fin}}\frac{V(\psi)}{V'(\psi)}\,d\psi = \frac{\psi_{\it fin}- \psi_{\it in}}{k_{(3)}},
\end{equation}
the value of the exponent $k_{(3)}$ is taken from (\ref{58}); $\psi_{\it in}$ and $\psi_{\it fin}$ are the values of the radion (inflaton) field in the beginning and in the end of the inflation. Thus e.g. for the value of $k_{(3)}=0,21$ (\ref{59}) it follows from (\ref{60}) that necessary number of $e$-foldings is reached if $\psi_{\it fin}- \psi_{\it in} > 20$.

The end of inflation where reheating begins is expected at the beginning of steep slope of the radion potential. Fig. \ref{AltFig1}, \ref{AltFig2} show that this happens somewhere at $\psi_{\it fin} > -20$. Hence initial position of the "lid" brane must be sufficiently deep in the throat: approximately $\psi_{\it in} < -40$.

We shall not come now into more detailed discussions on the capability of the proposed model to describe the slow-roll inflation consistent with increasing demands of the astrophysical observations \cite{WMAP}. Let us just look at the validity of the inequality $\psi_{\it in}<-40$ from the point of view of applicability of the low-energy string approximation. The permitted values of the isotropic coordinate $r$ must obey $r> r_{\it min}$ where $r_{\it min}$ is given in (\ref{14}). Corresponding minimal value $\psi_{\it min}$ is received from (\ref{45}) (where it is taken $y_{\it min}= \rho_{\it min}/L=r_{\it min}/L$). If we express $ML$ in (\ref{14}) from (\ref{26}) through the free parameter $g$ (\ref{27}) of the flux brane solution then $\psi_{\it min}$ and also the value of dimensionless radion potential (\ref{57}) at this point are found. For the type $IIA$ supergravity compactified to (3+1) dimensions this gives ((\ref{45}), (\ref{59}) with account of (\ref{14}), (\ref{26}) were used):

\begin{equation}
\label{61}
\psi_{\it min}=-168-152\cdot \ln g, \qquad V_{{\it D}(3)}(\psi_{\it min})=10^{15}\cdot g^{32}.
\end{equation}
For $g=1$ value of $\psi_{\it min}$ is essentially below the value demanded by inflation ($\psi_{\it in}<-40$). Of course it is possible to put $g \le e^{-1}$ in (\ref{61}) which will make $|\psi_{\it min}|$ of order one. But according to (\ref{26}) this will violate the basic inequality (\ref{15}) and hence invalidate all the approach. The choice of $g$ determines many predictions of the theory (see discussion in Subsec. {\it 4-b}).

The consistency of the theory demands also to make sure that effective radion potential (\ref{57}) which grows exponentially down the throat does not exceed Planck density. For the constant asymptotic of potential in the Type $IIB$ supergravity it was demonstrated in (\ref{53}). To check up this consistency condition in the Type $IIA$ supergravity it is necessary to know the value of the characteristic mass $\mu_{(3)}$ in (\ref{57}) (see below in Subsec {\it 4-c}).

\section{Estimation of values of mass scale \\
hierarchy and Dark Energy density}

{\large\it 4-a. A variable mass scale of matter in the early Universe.}

\vspace{0,5cm}

Following the conventional Randall and Sundrum approach \cite{Randall} we suppose that mass parameters of matter action written in the primordial metric of the action (\ref{1}) are of the fundamental scale $M$, e.g. action of scalar matter field $\Psi$ is given by:

\begin{equation}
\label{62}
S_{\it matter} = \int \left[-(1/2)(\nabla \Psi)^{2}-(1/2)M^{2}\Psi^{2}\right]\sqrt{-g^{(D)}}\,d^{D}x.
\end{equation}

Another conventional assumption says that matter is trapped at the "visible" IR brane, in our case at the IR end, $r=r_{\it IR}$, of the warped throat of the fluxbrane solution (\ref{7}). As it was said in the Introduction the necessity of "trapping" is questionable since massive matter experiences gravitational attraction to the IR end of the strongly warped space-time and may concentrate there even in absence of the "trapping" brane \cite{Katherine}. 

Anyway it is supposed that Standard Model resides at the IR end, hence integral over coordinate $r$ in the matter action ($\ref{62}$) is concentrated there. With this assumption mass of the visible matter is decreased as compared to that in (\ref{62}) by the value of warp factor $H^{\beta}$ in the metric (\ref{7}) at $r=r_{\it IR}$:

\begin{equation}
\label{63}
M \to M\,H^{\beta}(r_{\it IR}).
\end{equation}

This however is not the end of the story: it is necessary to write down the effective matter action in lower dimensions in the Einstein-frame metric $g_{\mu\nu}$ introduced in (\ref{40}) for manifold $M_{(p+1)}$ and in (\ref{55}) for (3+1) dimensions. We shall put down the final expression for the matter mass scale $m(\rho)$ depending on the radion field in (3+1) dimensions in general case when $p>3$ in the fluxbrane solution (\ref{7})-(\ref{9}). From (\ref{63}) and (\ref{55}) it follows:

\begin{equation}
\label{64}
m(\rho)=MH^{\beta}(r_{\it IR})\frac{M_{\rm Pl}}{[\Phi_{(3)}(\rho)]^{1/2}},
\end{equation}
Brans-Dicke field $\Phi_{(3)}(\rho)$ is given in (\ref{35}) where in case $p>3$ the additional factor $T_{(p-3)}$ (volume of compact extra space $K_{(p-3)}$ (\ref{8})) must be included in the RHS. Taking it into account the following expression for the mass scale hierarchy is received from (\ref{64}) and (\ref{35}):

\begin{equation}
\label{65}
\frac{m(\rho)}{M_{\rm Pl}}=\frac{H^{\beta}(r_{\it IR})}{\left[2(T_{(p-3)}M^{p-3})(ML)^{n+1}\Omega_{n}f(\rho/L)\right]^{1/2}}.
\end{equation}

From definition of function $f(y)$ (\ref{35}) and asymptotic (\ref{45}) follows exponential dependence of $m$ on the canonical radion field $\psi$ deep in the throat (for the Type $IIA$ supergravity dependence is the same as the asymptotic (\ref{59}) of the effective radion potential). Thus while radion field is rolling down the radion potential - at the supposed stages of inflation and reheating at the steep slope - mass scale of matter is decreasing. It is a task for future to study how this behavior may influence the observed picture of the Universe.

Mass scale of matter is stabilized at the modern observed value when radion field reaches the minimum of the radion potential at $\psi=0$.

\vspace{0,5cm}
{\large\it 4-b. Estimation of mass scale hierarchy in modern epoch.}

\vspace{0,5cm}

Thus mass scale hierarchy in modern epoch is given by expression (\ref{65}) where $\rho=r_{0}$ is to be taken, and $r_{0}$ is given in (\ref{23}).

We shall perform calculations for the Type $IIA$ supergravity (theory {\it C} in (\ref{49})) compactified to (3+1) dimensions. With account $r_{\it IR} \ll L$ in the argument of $H$ and expression (\ref{26}) for $(ML)$ it follows from (\ref{65}) for this theory:

\begin{equation}
\label{66}
\frac{m(r_{0})}{M_{\rm Pl}}=\frac{(r_{\it IR}/L)^{9/16}}{\left[2(TM)(ML)^{5}\Omega_{4}f(r_{0}/L)\right]^{1/2}}=1,2\cdot 10^{-3}\,\frac{(r_{\it IR}/L)^{9/16}}{\left[(TM)g^{5}\right]^{1/2}}.
\end{equation}
Value of mass scale hierarchy is determined by three parameters: location $r_{\it IR}$ of the IR end of the throat, period $T$ of the additional torus in (\ref{8}) ($K_{(p-3)}$ is $S^{1}$ for $p=4$) and dimensionless parameter $g$ (\ref{27}) of the elementary fluxbrane solution (\ref{7}). To receive the numbers it is necessary to set forth certain hypothesis:

\vspace{0,5cm}
{\it Hyp. 1.} Let us suppose, as it was suggested in \cite[d]{Altsh} and discussed in the Introduction, that visible matter is concentrated in the throat at the boundary of validity of the low-energy string approximation i.e. at $r=r_{\it min}$ given in (\ref{14}). This would mean $r_{\it IR}=r_{\it min}$ in (\ref{66}), and from (\ref{14}), (\ref{26}) for the Type $IIA$ supergravity it follows:

\begin{equation}
\label{67}
\frac{r_{\it IR}}{L}=\frac{r_{\it min}}{L}=4,1\cdot 10^{-16}\,g^{-32}.
\end{equation}

\vspace{0,5cm}
{\it Hyp. 2.} Let us use the Euclidian "time" version of the Schwarzshild type non-extremal modification (\ref{16}) of the elementary 4-brane solution as a tool to reduce from 4 to 3 space dimensions, and let us suppose that it provides smooth IR end of the throat. Then in expression (\ref{18}) for the period $T_{\theta}$ of the "time" torus we must put $r_{\it Sch}=r_{\it IR}=r_{\it min}$ ({\it Hyp. 1}) and put deficit angle coefficient $\delta_{d}=1$. Again with account of expression (\ref{26}) for $(ML)$ in case of the Type $IIA$ supergravity it follows from (\ref{18}) that:

\begin{equation}
\label{68}
T_{\theta}M=2,2 \cdot 10^{9}\,g^{17}.
\end{equation}
For $M=M_{\rm Pl}$ and $g=1$ we have $T_{\theta}=10^{-24}sm.$

Substitution of (\ref{67}), (\ref{68}) into (\ref{66}) gives:

\begin{equation}
\label{69}
\frac{m(r_{0})}{M_{\rm Pl}}=5,7 \cdot 10^{-15}\,g^{-29},
\end{equation}
which for $g=1$ approximately coincides with the observed ratio of the electro-weak and gravitational scales. This however can not be considered a great victory because of strong dependence of the result on free parameter $g$ (\ref{27}).

\vspace{0,5cm}
{\it Hyp. 3.} Let us consider in detail the third hypothesis: $g=1$. According to definitions of $g$ and $\sigma$ in (\ref{27}) and (\ref{2}) dimensionless parameter $g$ depends on the UV brane's tension $T_{\it br}$ and on the value of dilaton field at infinity, $\varphi_{\infty}$. It is natural to suppose that tension of co-dimension one brane in the action (\ref{1}) is determined by the fundamental string scale; then from (\ref{2}) follows $\sigma=M$ and (\ref{27}) gives:

\begin{equation}
\label{70}
g=e^{\frac{\alpha}{2(n-1)}\varphi_{\infty}}.
\end{equation}

Now two contradictory conditions of validity of the low-energy string approximation come into play. The first one (\ref{12}) demand $g<1$, whereas inequality (\ref{15}) ($ML \gg 1$) with account of (\ref{26}) says that $g$ must be of order of one or above. Thus $g=1$ ($\varphi_{\infty}=0$) is a natural choice, although perhaps not sufficiently well motivated.

\vspace{0,5cm}
{\large\it 4-c. Consistency of the value of the radion effective potential.}

\vspace{0,5cm}

Hypothesis formulated in the previous subsection permit to calculate in the Type $IIA$ supergravity the value of radion effective potential at the bottom (\ref{67}) of the throat and to compare it with Planck density. For the Type $IIA$ supergravity the following value of the characteristic density $\mu_{(3)}^{4}$ defined in (\ref{57}) is found with use of expressions (\ref{26}) for $(ML)$ and (\ref{68}) for $(TM)$:

\begin{equation}
\label{71}
\mu_{(3)}^{4}=1,1 \cdot 10^{-18}\,M_{\rm Pl}^{4}\,g^{-24}.
\end{equation}
The value of dimensionless radion potential $V_{(3)}(\psi_{\it min})$ (also defined in (\ref{57})) is given in (\ref{61}). Thus from (\ref{61}) and (\ref{71}) it follows:

\begin{equation}
\label{72}
\frac{\mu_{(3)}^{4}V_{(3)}(\psi_{\it min})}{M_{\rm Pl}^{4}}=1,1 \cdot 10^{-3} \, g^{8}.
\end{equation}
It is seen that the choice $g=1$ complies with the low-energy approximation demand: radion potential is below Planck density even at the point where curvature reaches the fundamental string scale.

\vspace{0,5cm}
{\large\it 4-d. Estimation of the Dark Energy density.}

\vspace{0,5cm}

It was shown in previous subsections that compactification from 4 to 3 space dimensions with a tool of the non-extremal bulk metric (\ref{16}) is crucially important for calculation of mass scale hierarchy in the considered approach. But introduction of this Euclidian "time" version of the Schwarzshild type modification of the bulk extremal fluxbrane solution inevitably violates junction conditions at the "isotropic" UV brane, as it is seen from discrepancy of expressions (\ref{28}), (\ref{29}). Since it is supposed that $r_{\it Sch} \ll L$ in (\ref{17}) (finally we take $r_{\it Sch}=r_{\it IR}=r_{\it min}$, {\it Hyp. 1, 2} in Subsec. {\it 4-b}) the discrepancy is quite small. To heal it the small anisotropy of the brane's energy-momentum tensor may be introduced, or bulk solution may be modified by introduction of small positive curvature in 4 dimensions, as it was done in \cite[a]{Altsh} where 6D model was considered. In what follows we just make a rude estimation of the possible corresponding deviation of the radion effective potential (\ref{57}) from its zero value at the minimum.

From (\ref{28}), (\ref{29}) it is possible to estimate the absolute value $|\delta\sigma|$ of the anisotropic variation of brane's characteristic mass $\sigma$ which can make junction conditions (\ref{28}), (\ref{29}) consistent:

\begin{equation}
\label{73}
|\delta\sigma| \cong U'(r_{0}) \approx L^{-1}\left(\frac{r_{\it Sch}}{L}\right)^{n-1},
\end{equation}
here $U(r)$ is given in (\ref{17}) and is taken at the brane's position $r=r_{0}$ (\ref{23}), prime means derivative over $r$, and we omitted all terms of order one like dimensionalities, $r_{0}/L$ etc.

Corresponding variation of the auxiliary potential ${\widetilde V}(\rho)$ at $\rho=r_{0}$ is immediately found from (\ref{37}); the same is the variation of the auxiliary potential ${\widetilde V}_{(3)}(r_{0})$ in the RHS of (\ref{57}) (two potentials differ only by factor $T_{(p-3)}$, see Subsec. {\it 3-c}). Finally the expected variation of the physical radion potential (\ref{57}) in the action (\ref{56}) in its minimum $\psi=0$ ($\rho=r_{0}$) is:

\begin{eqnarray}
\label{74}
&&|\delta [\mu_{(3)}^{4}V_{(3)}(0)]|=M_{\rm Pl}^{4}\frac{(|\delta\sigma|  L)e^{\gamma\varphi(r_{0})}H^{\xi}(r_{0})\,(r_{0}/L)^{n}}{4(ML)^{n+3}[f(r_{0}/L)]^{2}\Omega_{n}(T_{(p-3)}M^{p-3})}  \nonumber
\\
&&\approx \mu_{(3)}^{4}\left(\frac{r_{\it Sch}}{L}\right)^{n-1},
\end{eqnarray}
$\mu_{(3)}^{4}$ is defined in (\ref{57}); in the last aproximate expression we again omitted all terms of order one and used (\ref{73}) for $|\delta\sigma|$.

If we put in (\ref{74}) $r_{\it Sch}=r_{\it min}$ (\ref{14}) and for the Type $IIA$ supergravity compactified to 3 space dimensions ($D=10,\, n=4,\, p=4,\, \alpha=1/2$) use in (\ref{74}) the values (\ref{67}) and (\ref{71}) for $r_{\it Sch}/L$ and $\mu_{(3)}^{4}$ then it finally comes out:

\begin{equation}
\label{75}
|\delta [\mu_{(3)}^{4}V_{(3)}(0)]|= 8 \cdot 10^{-65}\,M_{\rm Pl}^{4}\,g^{-120}.
\end{equation}

For the "natural" choice $g=1$ ({\it Hyp. 3} in Subsec. {\it 4-b}) this value of the Dark Energy density is about 60 orders above the observed one \cite{Copeland}). Reissner-Nordstrom type generalization of the extremal $p$-brane solution may essentially improve the result. This was shown in \cite[a]{Altsh} for D6 version of the RS-model.

It is necessary to outline again that all the considerations of this subsection are essentially preliminary. In particular the sign of the estimated absolute value (\ref{75}) of the deviation of the radion effective potential is not established. Further work is needed to clarify the situation.

\section{Discussion}

\qquad Let us sum up some questions that need to be studied.

The idea that higher-dimensional magnetic monopole $p$-brane space-time may be quantumly born at the brink of applicability of the low-energy string theory description (i.e. at the surface $r=r_{\it min}$ (\ref{14})) does not look unrealistic. The characteristic length (\ref{10}) of the "newborn" should satisfy the strong inequality (\ref{15}). And it is necessary to suppose that together with magnetic $p$-brane responsible for appearance of the throat-like bulk solution (\ref{7}) the co-dimension one "heavy lid" UV  brane which cut the volume of extra dimensions was also born sufficiently deep in the throat (see discusion in the end of Sec. 3). Characteristics of two branes: magnetic charge $Q$ of the $p$-brane and tension $T_{\it br}$ (or mass $\sigma$ defined in (\ref{2})) of the co-dimension one brane must be "fine-tuned" like in (\ref{25}).

After these vague quantum preliminaries, which deserve better clarification, early evolution of the Universe may be described by the effective action (\ref{56}) where position of the co-dimension one "lid" brane is given by the canonical radion (inflaton) field $\psi$. This dynamics governs brane's slow rolling down the radion effective potential (UV brane moves up the throat) and simultaneous inflation of the Universe in 3 space dimensions. When UV brane reaches the steep slope close to the top of the throat (as it is seen at Fig. \ref{AltFig1}, \ref{AltFig2} in Subsec. {\it 3-b}) reheating begins. After UV brane is stabilized in the minimum of the radion effective 
potential the conventional evolution of the Universe takes place. The task for future is to study if this scenario can model the simplest class of inflationary theories \cite{Guth} capable to describe the plethora of astrophysical observations \cite{WMAP}.

Randall and Sundrum tool \cite{Randall} of calculation of the electro-weak hierarchy may be applied in this scenario, and perhaps may give a satisfactory result (see (\ref{69})) if it is supposed that observed massive matter, including ourselves, remains concentrated at the very bottom of the throat where evolution began after quantum tunneling "from nothing". This {\it Hyp. 1} in Subsec. {\it 4-b} looks strange and it would be quite interesting to find some grounds for it.

The most promising in the proposed approach is perhaps the magnetic 4-brane solution in the Type $IIA$ supergravity ($D=10$, $p=4$, $n=4$, $\alpha=1/2$ in the action (\ref{1}) and in the bulk solution (\ref{7})). This model demands additional compactification from 4 to 3 space dimensions. The subtleties of this compactification performed with a tool of the Euclidian "time" version of the non-extremal Schwarzshild type modification (\ref{16}) of the extremal solution (\ref{7}) permit to fix the additional modulus (period of the "time" torus) and perhaps even open the way for calculating the Dark Energy density (Subsec. {\it 4-d}). This preliminary trend of thought needs further development.

Many other questions need to be clarified. Some of them are:

- In calculating radion effective potential (Sec. 3) we ignored the lower limit of integration over isotropic coordinate $r$. This point must be studied better together with description of the IR end of the throat.

- It would be interesting to determine the characteristic time of the runaway decompactification of the extra dimensions by tunneling through the barrier at $\psi>0$ seen at Fig. \ref{AltFig1}, Subsec. {\it 3-b}.

- We did not discuss the role of quantum corrections and supersymmetry.

- And with primitive compactification on $S^{n}$ in the elementary fluxbrane solution (\ref{7}) it is hardly possible to describe the Standard Model world.

\section*{Acknowledgements} Author is grateful for plural discussions to participants of the Quantum Field Theory Seminar of the Theoretical Physics Department, Lebedev Physical Institute, and is obliged to R.K. Galimov, D.V. Nesterov and M.O. Ptitsyn for assistance. This work was partially supported by the grant LSS-4401.2006.2

\section*{Appendix: Nonequivalence of the electric \\
fluxbrane and magnetic fluxbrane solutions \\
in the effective action calculations}

\vspace{0,8cm}

\qquad Consider action (\ref{1}) with arbitrary $q$-form ($q=n$ in case space-time (\ref{7}) describes magnetic fluxbrane solution and $q=p+2$ when dual electric fluxbrane solution is considered). Lagrangian of the action (\ref{1}) calculated at any solution of the dynamical equations following from this action is given by:

\begin{equation}
\label{76}
L=-\left(\frac{q-1}{D-2}\right)\frac{1}{q!}e^{\alpha\varphi}F_{(q)}^2+\frac{1}{D-2}\sigma e^{\gamma\varphi}\delta^{(1)}\frac{\sqrt{-h^{(D-1)}}}{\sqrt{-g^{(D)}}}.
\end{equation}

General consistency formula (12) of the paper \cite{Leblond} written down when arbitrary parameter $\alpha_{[27]}$ of this paper is put equal to $p$ ($p$ is number of the uncompactified space dimensions like in the elementary fluxbrane solution (\ref{7}) above) and for zero curvature of the manifold $M_{(p+1)}$ says that the combination

\begin{equation}
\label{77}
(p+1)T_{m}^{m}-(D-p-3)T_{\mu}^{\mu}
\end{equation} 
of traces of the energy momentum tensor in compact space - $T_{m}^{m}$ (here index $m$ embraces all, i.e. $y^{i}$ and $r$, extra coordinates in (\ref{7})), and in the uncompactified space-time - $T_{\mu}^{\mu}$ being weighted
with a volume-factor of the extra dimensions is a full divergence in extra dimensions. Hence integral of the combination (\ref{77}) over compact extra space is equal to zero. This is a version of the famous consistency conditions. 

Thus action (\ref{1}) calculated at the solution of the dynamical equations and integrated over extra dimensions (which gives the value of the effective radion potential in its extremum at $\rho=r_{0}$ (\ref{23})) is equal to zero if Lagrangian (\ref{76}) is proportional to the combination (\ref{77}).
Let us show that this is the case for magnetic fluxbrane solution and not for dual electric one. 

Substitution in (\ref{77}) of the energy-momentum tensor of the dilaton, $q$-form and brane terms in (\ref{1}) (where $F_{n}$ is replaced by $F_{q}$) gives:

\begin{eqnarray}
\label{78}
&&(p+1)T_{m}^{m}-(D-p-3)T_{\mu}^{\mu}=\frac{e^{\alpha\varphi}}{2(q-1)!}\left[-(D-p-3)F_{(q)\mu...}F_{(q)}^{\mu...}\right. \nonumber \\
&&+\left.(p+1)F_{(q)m...}F_{(q)}^{m...}-\frac{p+1}{q}F_{(q)}^{2}\right]-\frac{p+1}{2}\sigma e^{\gamma\varphi}\delta^{(1)}\frac{\sqrt{-h^{(D-1)}}}{\sqrt{-g^{(D)}}}.
\end{eqnarray} 

Now it is easily seen that expression (\ref{78}) is proportional to Lagrangian (\ref{76}) when $q$-form field strength "lives" only in extra dimensions, i.e. when $F_{(q)\mu...}=0$ and $F_{(q)m...}F_{(q)}^{m...}=F_{(q)}^{2}$. This is true for the magnetic monopole solution in (\ref{7}) and is not true when electric fluxbrane solution with dual $(p+2)$-form is considered.

Direct calculation of the action (\ref{1}) on the electric fluxbrane background when brane is located at the position (\ref{23}) fixed by junction conditions also gives non-zero value; this results in unphysical vacuum energy in low dimensions and violates the consistency demands.

\end{document}